\documentclass[10pt,twocolumn,floatfix,reprint]{revtex4}

\usepackage{graphicx,amssymb,amsmath,hyperref}
\usepackage{xfrac}
\usepackage{courier}

\begin{document}

\title{Statistical analysis of chess games:\\
  space control and tipping points}

\author{Marc Barthelemy$^{1,2}$}

\affiliation{$^1$ Universit\'e Paris-Saclay, CNRS, CEA, Institut de Physique Th\'eorique, 91191, Gif-sur-Yvette, France}

\affiliation{$^2$ Centre d’Analyse et de Math\'ematique Sociales (CNRS/EHESS) 54 Avenue de Raspail, 75006 Paris, France}

\begin{abstract}

  Moves and games are usually analyzed in detail by professional players but very few  statistical analysis of the game are available. Here, we analyze moves in chess games from a statistical point of view. We first focus on spatial properties and the location of pieces. We first show - as intuitively expected - that the number of possible moves during a game is correlated with its outcome. We then study heatmaps of pieces (for a given opening), and we show that the spatial distribution of pieces varies less between human players than with engines (such as Stockfish): engines seem to use pieces in a very different way as human did for centuries. These heatmaps also allow us to construct a distance between players that characterizes how they use their pieces. In a second part, we focus on the best move and the second best move found by Stockfish and study the difference $\Delta$ of their evaluation. We found different regimes during a chess game. In a `quiet' regime, the engine evaluation for the best and second best moves are not very different ($\Delta$ small), indicating that many paths are possible for both players. In contrast, there are also `volatile' regimes characterized by a `tipping point', for which $\Delta$ becomes large and display large fluctuations. Furthermore, we found that for a large number of games, the distribution of $\Delta$ can be fitted by a power law $P(\Delta)\sim \Delta^{-\beta}$  with an exponent that seems to be universal (for human players and engines) and around $\beta\approx 1.8$. The existence of such a power law indicates that during chess games, we can have regimes with very large fluctuations of $\Delta$ indicating a relatively high frequency of tipping points that determine in a large part the outcome of the game. Finally, we conclude by mentioning possible directions of research for a quantitative understanding of chess games such as the structure of the pawn chain, the interaction graph between pieces, or a quantitative definition of critical points.
  
\end{abstract}

\maketitle

\section{Introduction}

The game of chess always provided an interesting playground for testing computer capabilities and algorithmic advances \cite{Levy:1988,Silver:2018}. This is in particular due to the fact that the number of games is too large to be exhausted by naive search . The mathematician Shannon discussed some properties of chess \cite{Shannon:1950}, including an estimate for the number ${\cal N}$ of sensible games that he found to be of the order of $10^{120}$. This conservative estimate was based on the fact that there are approximately $30$ possible (legal) half moves (i.e. a move for each player) for each player and that a typical game lasts about $40$ moves (which represents $80$ half moves). This immediately  gives ${\cal N}\sim 30^{80}\sim 10^{120}$. Other results improved that estimate with upper and lower bound and in \cite{Steinerberger:2015} the upper bound was proposed ${\cal N}< 4\times 10^{37}$ in the absence of promotions (see \cite{Tromp:2022} for a detailed discussion of the number of games).

\begin{figure}[!htbp]
	\centering
        \includegraphics[width=0.4\textwidth]{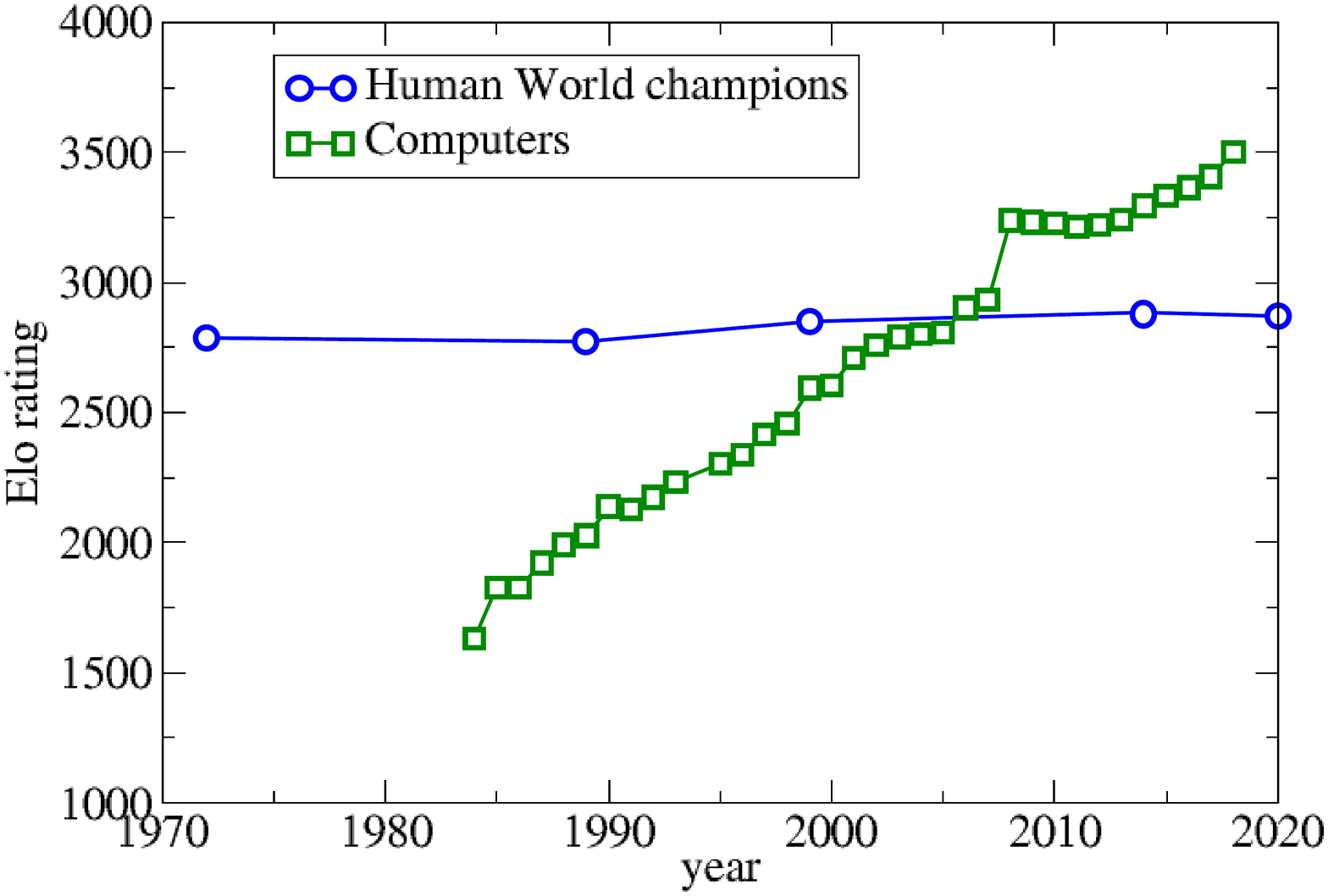}
        \vspace*{-3mm}
        \caption{Evolution of the Elo rating for human players and computers.}
        \vspace{-5mm}
	\label{fig:elorating}
\end{figure}

It is then not a surprise that chess engines include methods such as machine learning that allowed them to go far beyond human capabilities \cite{Sadler:2019}. In fact, computers and chess have a long history together \cite{Sadler:2019}, and already in 1950, Shannon published a paper \cite{Shannon:1950} about programming chess, and mathematicians such as Turing proposed simple algorithms to evaluate a position \cite{Levy:1988}. Computer competitions developed at a steady rate and in 1974 the first world computer championship took place in Stockholm (Sweden). By 1985 computers reached the level of $1800$ Elo corresponding to the average level of a chess club player. In 1997, the revenge match between the world chess champion Kasparov and `Deep-Blue' 
a chess-playing expert system run on a IBM supercomputer led to the first victory ($3\sfrac{1}{2}-2\sfrac{1}{2}$) of 
a computer against a reigning world champion. Although Deep Blue's evaluation function depends on a large number of parameters to be tuned, this victory is considered as a milestone in the history of artificial intelligence. Since then computers steadily defeated humans (see Fig.~\ref{fig:elorating}, and the best engines (Stockfish, Lc0, etc.) reach now Elo ratings close to $4000$ \cite{rating}, far above the current world champion Magnus Carlsen (with an Elo $\sim 2900$). We note here that many studies were devoted to the Elo rating and its evolution \cite{Simkin:2012, Fenner:2012,Alliot:2017}.
The combination of these engines and professional players allowed for a renewal in chess theory, refuting sharp opening variations and proposing new schemes (for a more detailed discussion, we refer the interested reader to \cite{Sadler:2019} where Alpha0's openings, moves, and games were analyzed in detail by professional players). One can however safely say that these engines are still in the `black box' regime and it is unclear how they apprehend strategical principles. As in other domains, explainability of the results is essential for users (particularly so for health applications) to effectively understand and manage powerful AI applications. In this respect, we believe that it is important to study the results found by machine learning means which might provide some clues on how these methods effectively reduces the complexity of the system and the data. Chess can then be seen as a good test bed for developing ideas useful for explainable AI \cite{Gunning:2019}.


From a purely theoretical point of view, a chess game can be
represented by a decision tree where the leaves denote the end of the game which can
be either a win, a loss or a draw. In this representation, the decision to make
a move is motivated by strategical arguments or tactics. In general, in the middlegame
when no tactics is involved the main goal is to find the move leading to a branch of the decision
tree which contains a favorable fraction of outcomes (characterized by a majority of leaves with
wins). This is a bit different in the endgame where calculation prevails and where a single path can
lead to victory and any deviation from it leads to a draw or a loss.

The question is then how to find the best move, in particular in the middlegame when
calculations are quasi impossible due to the explosive combinatorics. The need for a scientific approach to chess games was advocated by many players, and in particular Richard R\'eti, a famous player proponent of hypermodernism, proposed the idea of a scientific approach to chess. Nowadays, many ideas and principles are now common to many players such as the control of the center, the control of dark or white squares, the structure of pawns, etc., and are the first hints towards a scientific theory of chess. More recently, the grand master Iossif Dorfman \cite{Dorfman} in order to go beyond these usual strategical considerations, proposed some sort of algorithm. The main idea is that during the game there are some `critical points' that necessitate a careful assessment of the position, and depending on the assessment different types of moves are needed. From a theoretical point of view this method points to many interesting ideas including the existence of critical points (we discuss this further at the end of this paper). 

The explosion of available data, essentially through the advent of online chess platforms, made it possible to analyze a large number of games and to introduce some ideas coming from the statistical physics and more generally from complex systems \cite{Perotti:2013,DeMarzo:2022,Blasius:2009,Maslov:2009,Schai:2014,Schai:2016,Chow:2023} (in particular, see the recent paper \cite{DeMarzo:2022} and references therein). Scientists focused on the statistics of openings and found that the frequency of opening moves are distributed according to a power law, leading to a Zipf's law whose origin probably lies in the self-similar nature of the game tree \cite{Blasius:2009,Maslov:2009}. Other studies revealed long-range memory effects \cite{Schai:2014,Schai:2016} in chronological ordered chess databases that vary with the level of players. A recent interesting study \ref{Chow:2023} focused on gaming behaviors as a function of skills and explored the diversity in openings. Here, in contrast with these previous studies, we propose to focus on spatial aspects of chess and on the best move properties, in a statistical perspective. Averaging over a large set of games, we will focus on the spatial distribution of pieces, and the existence of tipping points where a move can be decisive with respect to the game's outcome.

\section{Space on the chessboard}

\subsection{Number of legal moves}

The number of legal moves (authorized by the rule of the chess game) varies during the game (Fig.~\ref{fig:numbermove}), starting at $20$ for the first move, and typically increases before eventually decreasing (due to the decreasing number of pieces on the board).  We show in Fig.~\ref{fig:numbermove} the average number of legal moves versus the ply (also called `half move') for the World Rapid and Blitz Championships (Warsaw 2021, \cite{data:blitz}).
\begin{figure}[!htp]
	\centering
		\includegraphics[width=0.5\textwidth]{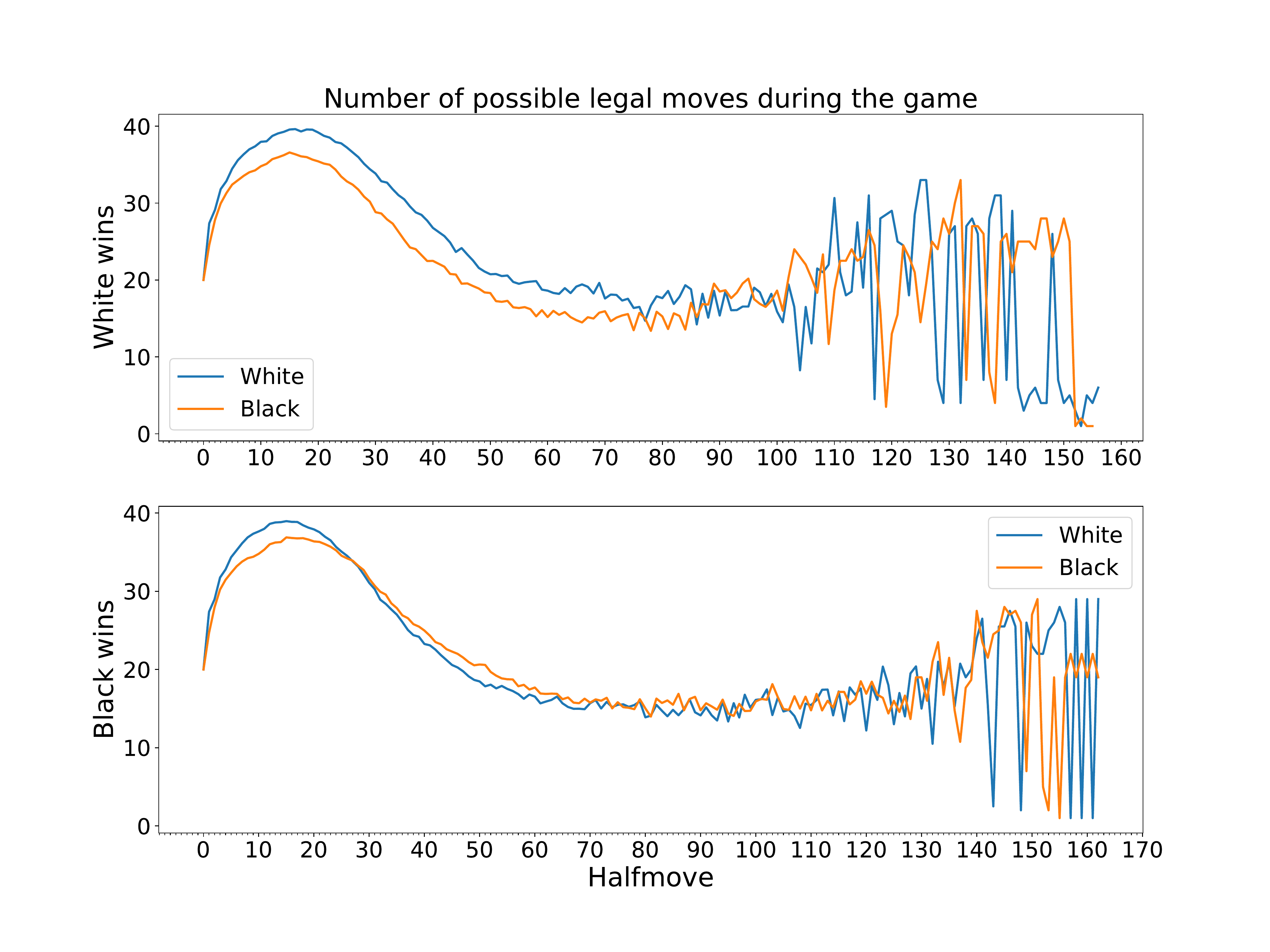}
	\caption{Average number of legal moves when white (top) or black (bottom) wins computed for $2977$  games played in the world rapid and blitz championship 2021 \cite{data:blitz}.}
	\label{fig:numbermove}
\end{figure}

The number of possible moves is a good indicator of how much space (measured here are the number of squares where one can move to) is controlled: the winner of the game ends up in general with a larger number of possible moves, due to a larger number of pieces or at a high level (when the number of pieces is in general the same) to a better spatial organization of its pieces. White always has the advantage at the beginning of the game, the goal for black is then to counter this advantage (and we observe that in blitz games this happens approximately at the 30th half move).

The number of possible moves at each round that varies at different stages of the game. The histogram of this number is shown in Fig.~\ref{fig:numbermove} for games between two of the best engines alpha0 and stockfish \cite{A0vsSF}. The average number of legal moves is of order $25$ (the small numbers observed in this histogram are due to check situations where moves are forced). What is however crucial is that this number is not constant throughout the game. Indeed,  this average number starts with $20$, reaches a peak and decreases when there are less and less pieces (see Fig.~\ref{fig:numbermove}). 
\begin{figure}[!ht]
	\begin{center}
		\includegraphics[width=0.45\textwidth]{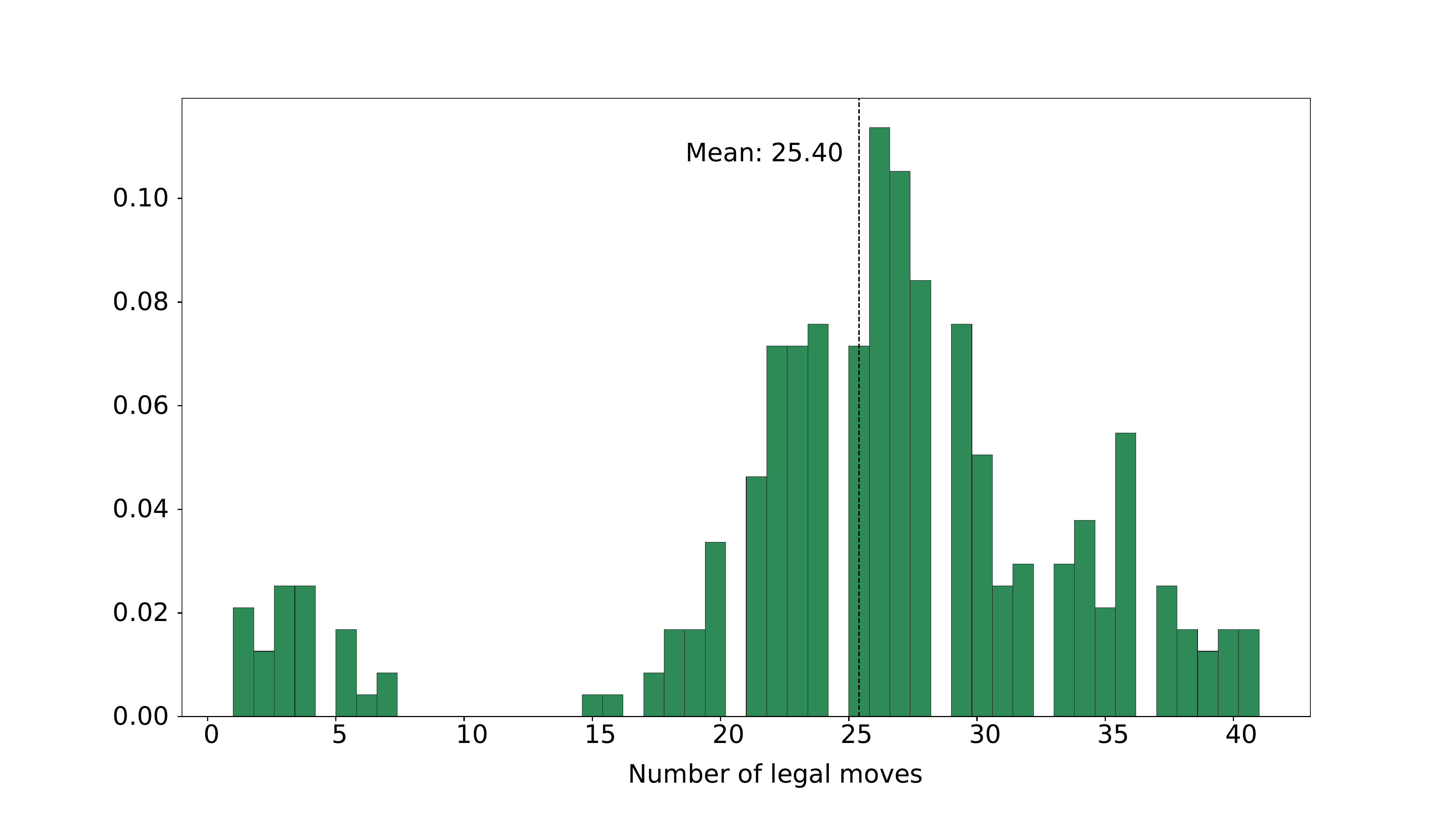}
		\includegraphics[width=0.45\textwidth]{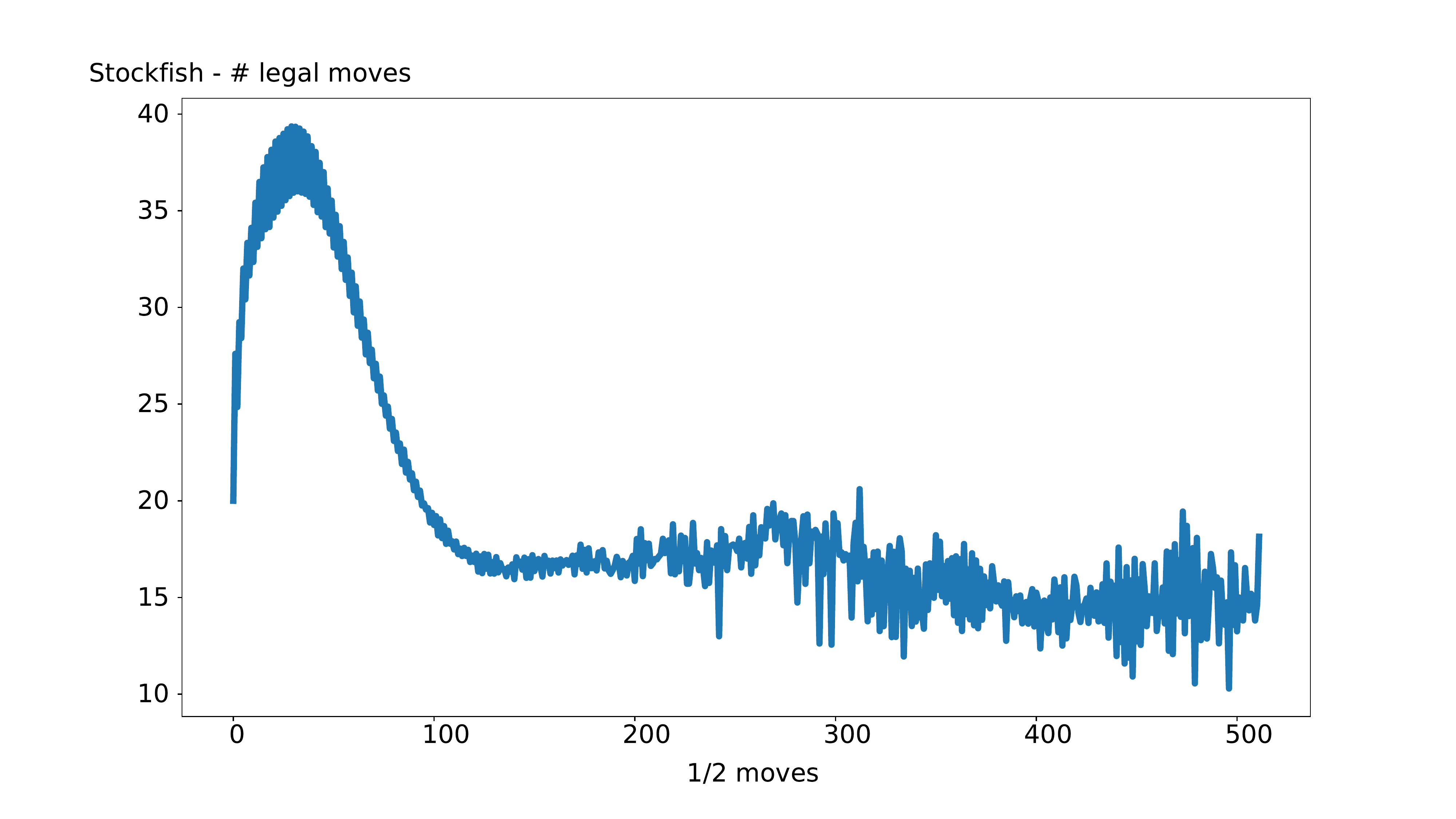}
	\end{center}
	\caption{(a) Histogram of the number of legal moves during games played between alpha zero and Stockfish. (b) Average number of legal moves during $3011$ games played by Stockfish \cite{data1}. We observe that this number varies from $20$ to $40$ in typically $20$ moves and then decreases from $40$ to $20$ also in $20$ moves.}
	\label{fig:numbermove}
\end{figure}
The average number of moves (average over many games) display a typical behavior shown in Fig.~\ref{fig:numbermove} (b). Typically the number of legal moves varies from $20$ to $40$ in $20$ moves and decreases from $40$ to $20$ in also $20$ moves. This information might be useful for finding an estimate for the number of possible games \cite{Tromp:2022}.

\subsection{An illustration}

\begin{figure}[!ht]
	\includegraphics[width=0.5\textwidth]{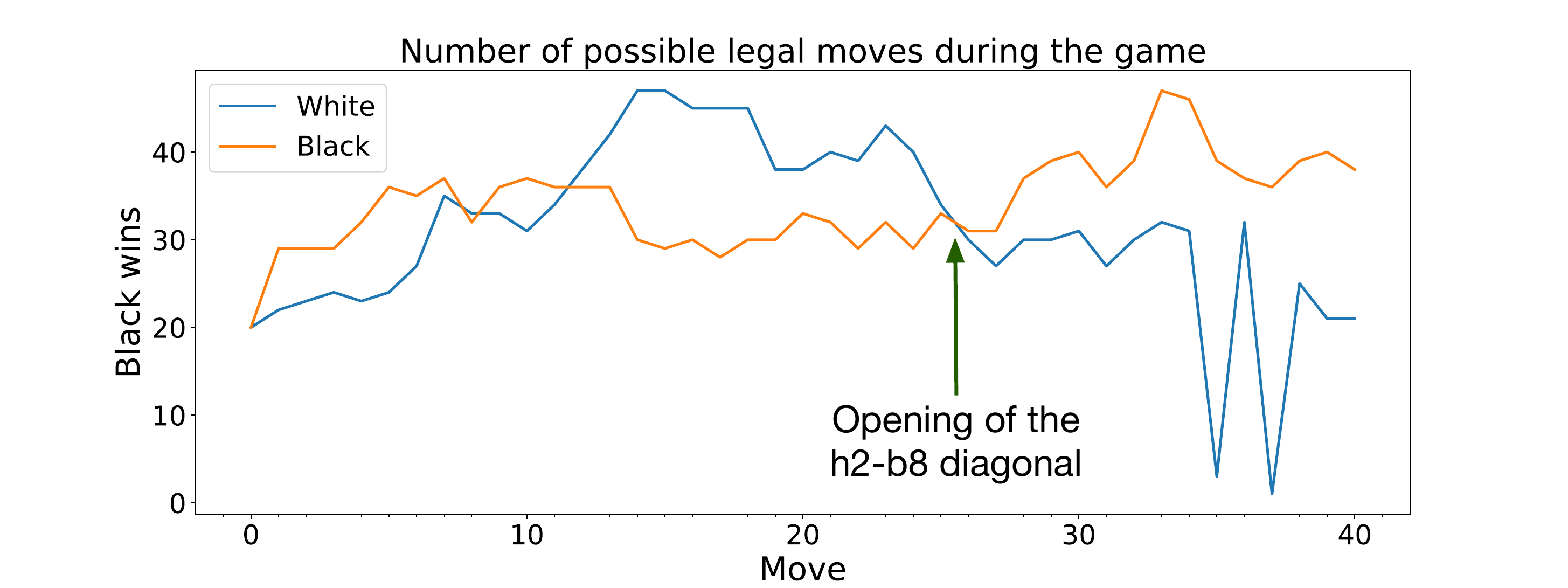}
	\caption{Example of the game Mchedlishvili-Van Forrest (Chennai Olympiad 2022). (Top) Evolution of the number of legal moves for both players during the game (blue curve for white and orange curve for black). A crucial move was done at move 25 for black (or equivalently at half move 50): the advantage in space becomes important for black (by opening a large diagonal) and the game is in favor for black.}
	\label{fig:vanforrest}
\end{figure}

The advantage of space is thus crucial and in general corresponds to a positive outcome. We illustrate this on the example of the game played between the grand masters Mchedlishvili and Van Forrest at the Chennai Chess Olympiad \cite{Chennai}. We plot the number of possible legal moves versus ply and show the result in Fig.~\ref{fig:vanforrest}.
From this plot, we observe that the turning point of the game seems to be at move $25$ when black played \texttt{Rd4}. The exchange of this rook after \texttt{Nxd4} and \texttt{exd4} sacrifices the quality but allows to open the diagonal \texttt{b8-h2} increasing in this way the number of squares available for black. After this move, white basically lost (as can be seen in the Stockfish evaluation) which was indeed the case.

\subsection{Spatial localization of pieces}

\begin{figure}[!h]
  \includegraphics[width=0.5\textwidth]{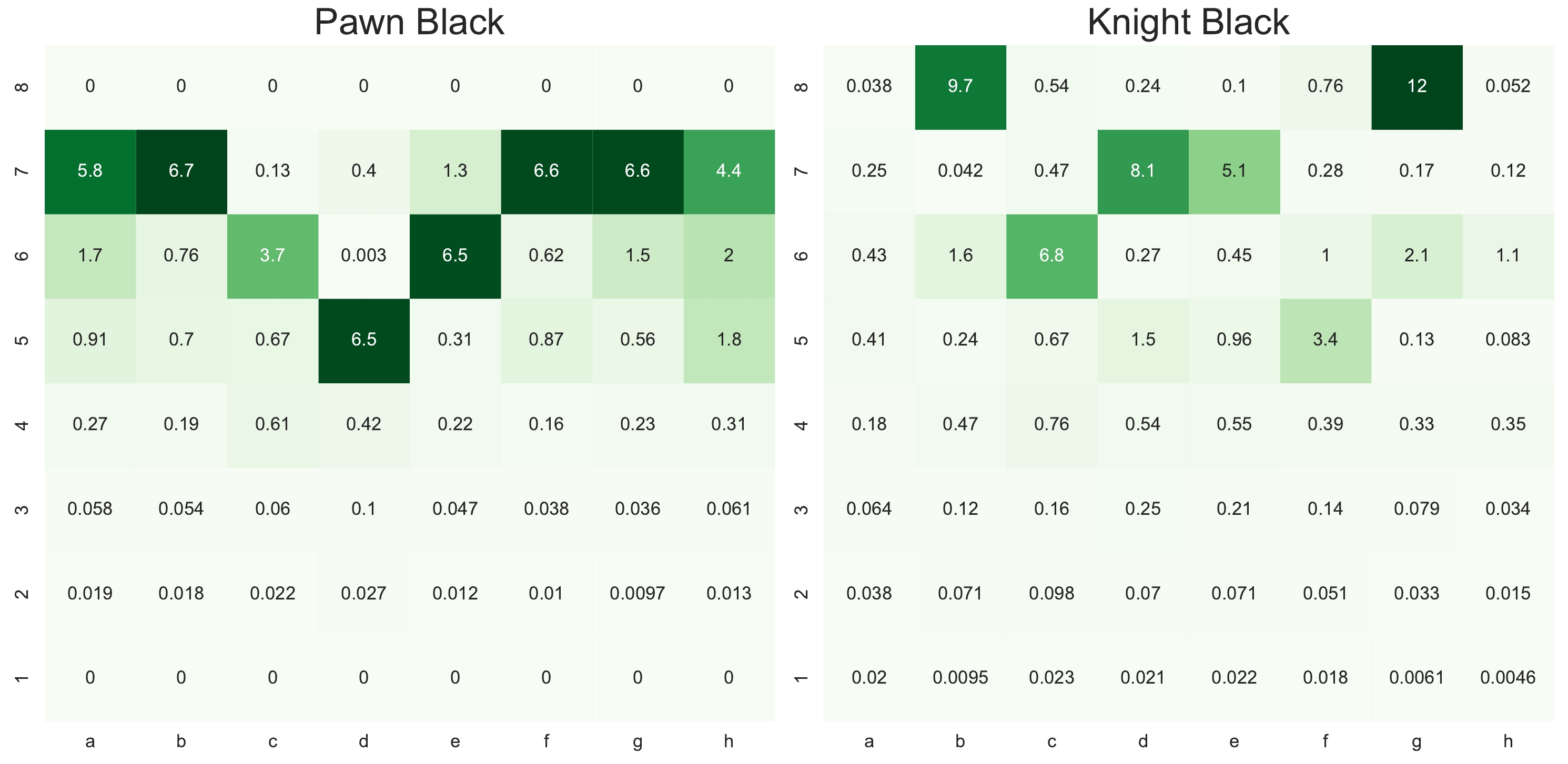}
	\caption{Heatmap: probability of presence for black pawns and knights in games with opening B12 (Caro-Kann). This heatmap were obtained for $35,909$ games \cite{data1}. The probabilities shown here  are normalized by $1/64$ which corresponds to the uniform distribution of pieces over the whole board (the color is darker for larger values of the normalized probability).}
	\label{fig:heatmap1}
\end{figure}

In order to understand statistically how the different players use their pieces, we computed the `heatmap of pieces' that represent the probability $p_i$  for a piece to be at a position $i$ for a large number of games. The position of pieces will in general strongly depend on the opening that governs the subsequent evolution of the pieces
and where they will be located (until the middlegame at least). In order to illustrate these results, we choose a variation of the Caro-Kann opening with ECO code B12 (Encyclopaedia of Chess Openings (ECO, see for example \cite{ECO}) which is characterized by \texttt{1.e4c6 2.d4}. The data by opening is available at \cite{data1} and for this opening we compute the heatmap of pieces averaged over $35,909$ games and show the result for black pawns and knights in Fig.~\ref{fig:heatmap1}
We observe on this heatmap some typical facts about this opening such as the pawn structure (with pawns usually located at \texttt{d5} and \texttt{e6})
and a knight usually on \texttt{d7} or \texttt{c6} and the other one staying a long time on \texttt{g8}. We also note the relatively frequent attack on the h-file with the pawn going to \text{h5}.

\begin{figure}[!h]
	\includegraphics[width=0.5\textwidth]{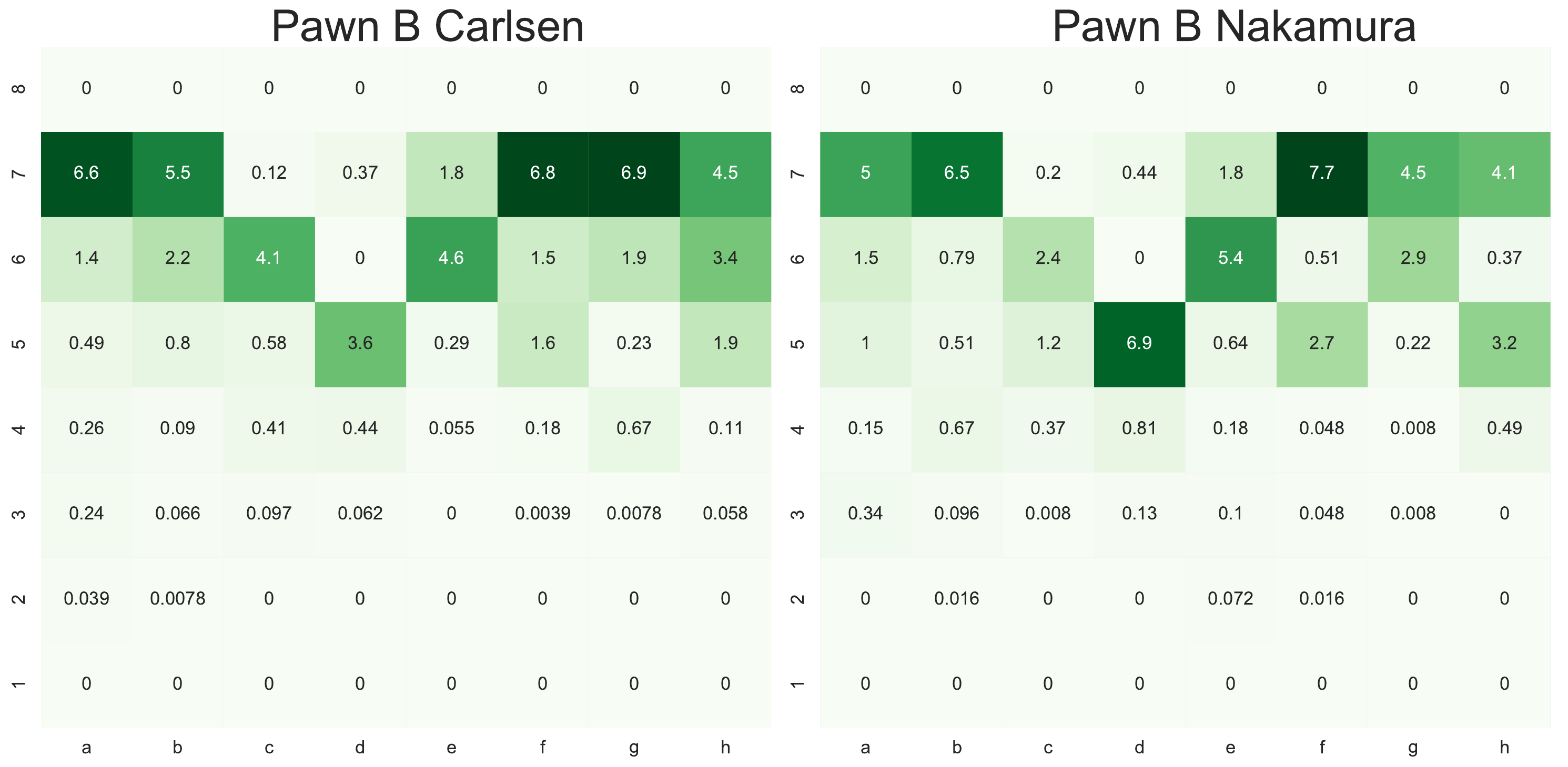}
	\caption{Heatmaps for pawns in the B10-19 openings (Caro-Kann and its variations) for Carlsen and Nakamura (data from \cite{data1}). The probabilities shown here are normalized by $1/64$ which corresponds to the uniform distribution of pieces over the whole board (the color is darker for larger values of the normalized probability).}
	\label{fig:heatmap2}
\end{figure}

We can also compare different players according to the frequent localization of their pieces for a given opening, which allows to construct some sort of distance between them. For example, we show in Fig.~\ref{fig:heatmap2} the heatmaps for black pawns obtained for openings B10-B19 (the Caro-Kann and its different variations) for Carlsen and Nakamura. Carlsen is the world champion (until 2023) and Nakamura is ranked fifth world player and despite all the analysis and modern `theory', we observe that for the same opening they use their pieces differently. We won't discuss these differences in detail but on the example of black pawns (ie. when these players use the Caro-Kann defense when playing black against \texttt{1.e4}), we observe for example that Carlsen likes to move its pawn on the \texttt{f}-file (on \texttt{f5} or \texttt{f6}) while Nakamura moves it less frequently but when he does it, it is preferentially on the \texttt{f5} square. In sharp contrast, Nakamura leaves its pawn on \texttt{d5} while Carlsen often exchanges it with enemy pawns. We see here on this particular example how a statistical approach could inform us about strategical elements. Of course these differences could result as a response of the opponent, typically here \texttt{3.e5} versus \texttt{3.Nc3}, but the main goal here is  to show that such a statistical approach contains potentially useful information.

\subsection{Wasserstein distances between players}

We can compare these heatmaps for all the pieces for two different players (for a given opening). We then use the `Earth moving distance' (also called 1-Wasserstein distance, see for example \cite{wiki:wasser,Wasser}). Intuitively, if each distribution is viewed as a unit amount of earth (soil) piled on an area, the metric is the minimum `cost' of turning one pile into the other, which is assumed to be the amount of earth that needs to be moved times the mean distance it has to be moved (which is why this metric is also known as the `earth mover's distance'). For 2d heatmaps as we have here, we consider the bipartite graph connecting squares on one heatmap to the squares on the other one and where the weight of edges is given by the chessboard distance (also called the Manhattan or taxicab distance). 

\begin{figure}[!ht]
		\includegraphics[width=0.5\textwidth]{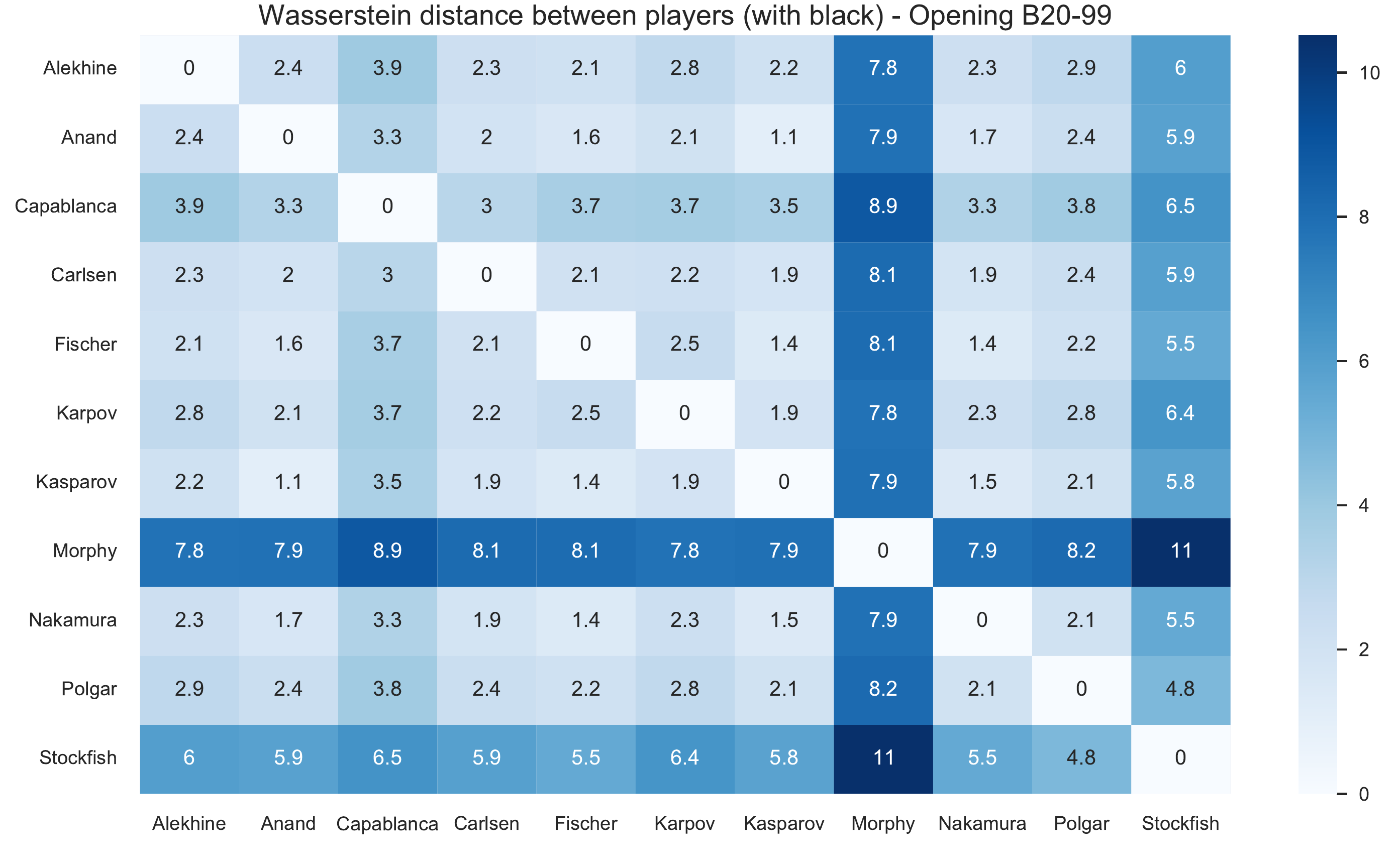}
	\caption{Statistical comparison of the spatial localization of black pieces between various world class champions and Stockfish15. The numbers (and the color: the darker and the larger the distance) indicate the Wasserstein distance between the players computed from the heatmaps with Eq..~\ref{eq:wasser}. }
	\label{fig:wasser}
\end{figure}

We thus compared several world-class players between them for a given opening. We also included here the the best engine so far, Stockfish 15 \cite{SF}, in order to compare it to the best human players. Here, we choose the example of the Sicilian opening (with ECO codes B20-B99) and which is characterized by the moves \texttt{1.e4c5}. For a pair $(\alpha,\beta)$ of players in this list, we compute the heatmaps for all black pieces $i\in\{p,n,b,r,q,k\}$ (and for $i\in\{P,N,B,R,Q,K\}$ in case we want to compare an opening played with white), and then compute the Wasserstein distance $d_W^i(\alpha,\beta)$ between these players for each piece $i$. We then construct the total distance between the players $\alpha,\beta$ by summing over all black pieces
\begin{align}
  d_W(\alpha,\beta)=\sum_{i\in\{p,n,b,r,q,k\}}d_W^i(\alpha,\beta)
  \label{eq:wasser}
\end{align}
We represent this Wasserstein distance on the Fig.~\ref{fig:wasser}.
The larger this distance between two players and the bigger the difference in using their pieces. For a given player, say Kasparov for example, we can then identify the other players with a similar game: Anand is the closest, followed by Nakamura. We also immediately see that a player such as Morphy (a player of the 19th century) played chess in a very different way as modern players, as expected. We can also see on this figure that Stockfish is very different from the other players (the closest player being Polgar, followed by Nakamura) indicating a very different use of the pieces.

\section{Tipping points}

\subsection{Best and second best move gap $\Delta$}

The idea in this section starts from the following question: during a game do we have the choice between different variations or is there a `forced' move that we have to make in order to avoid to be defeated ?  In order to answer this question, we look at the two best moves computed by the stockfish engine \cite{SF}.
The best move corresponds to a Stockfish evaluation $E_0$ and the second best move to $E_1$.
For white we have $E_0>E_1$ and for black $E_0<E_1$ (when the evaluation is positive, white wins and when negative black wins).
We  define the quantity
\begin{align}
  \Delta = E_0 - E_1
\end{align}
which characterizes the difference of quality between the best move and the second best move. When $\Delta$ is small, we have a choice and in the opposite case, the best move is much better than the second best move and we are in a forced move case.

\begin{figure}[!ht]
          \includegraphics[width=0.5\textwidth]{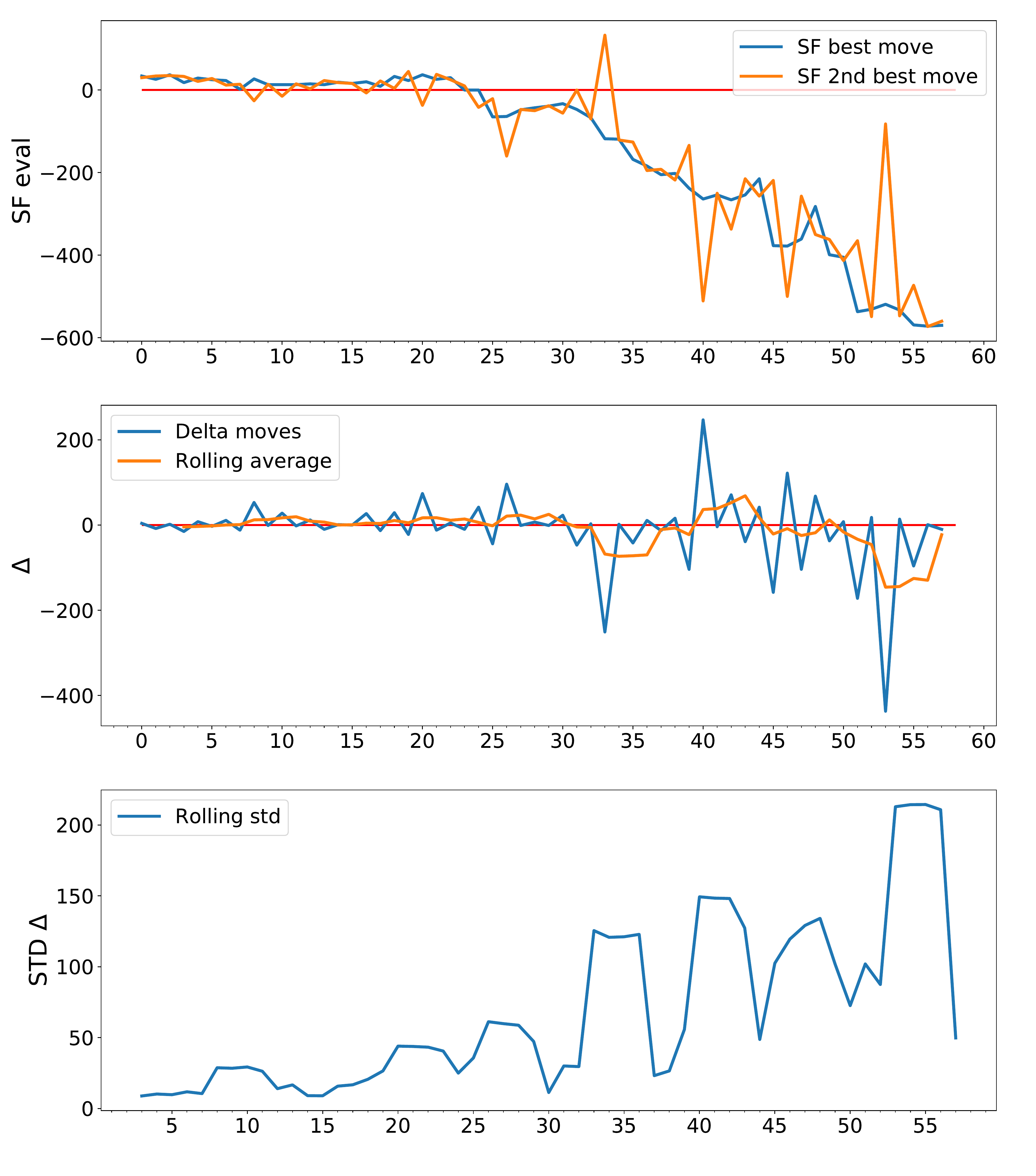}
	\caption{Calculation of $\Delta$ on the example of a game between Ding and Nepomniachtchi (Astana KAZ, rd 2, Apr-10 World championship 2023). a) Stockfish evaluation (in centipawn units) for the best move and the second best move. (b) Evolution of $\Delta=E_0-E_1$ during the game (together with its rolling average computed over 4 half moves). (c) Evolution of the Standard deviation (STD) of $\Delta$ during the game. } 
	\label{fig:delta}
        \vspace{-5mm}
 \end{figure}

We analyzed various games and computed this quantity $\Delta$ during their evolution. We observed that during a game there is in general a succession of different regimes (with variable length) with either small
 fluctuations of $\Delta$ or in contrast with large ones. Small values and fluctuations
 of $\Delta$ indicates that the best move and the second best move are almost equivalent. 
 We are then in the presence of possible different variations without a clear
 outcome. In contrast, in the `volatile' regime with large fluctuations of $\Delta$, it is
 critical to find the best move, as it is by far better than the second best move. This can
 happen for trivial reasons: if the opponent proposes an exchange and captures a piece, it is
 in general compulsory to capture it (if not, we loose a piece). However, there are more subtle
 cases where $\Delta$ is large but there is no immediate reason for distinguishing the best
 move and the second best one. This apparently happens in most games and corresponds
 then to a `tipping point' as the choice will mostly determine the outcome of the game. This
 for example happens in the Ding-Nepomniachtchi game (see Fig.~\ref{fig:delta}) at the $25^{th}$
 half move: the best move is \texttt{d4c5} (the Stockfish score is $19$ in centipawn units) that modifies the pawn structure,  and the second best move is \texttt{b2b3} (a pawn move with score $-58$). In this case we then have $\Delta=77$ (much larger than
 a typical value of order $10$). Finding the best move is then crucial here, and in this case Ding actually
 missed both of these moves (and played \texttt{e3e4} with a score equal to $-65$ !).
 
 The tipping points defined here thus separate a regime with small variance of $\Delta$ and a regime with a large variance. We thus plot the standard deviation of $\Delta$ (STD) and show the result in Fig.~\ref{fig:delta}. Again, in the `volatile' regime, we are in a dangerous zone
 and to make a mistake is almost fatal. It is interesting to note that the values of $\Delta$ for black and white
 are strongly correlated so that a regime is usually dangerous for both sides. We note that the length of these different regimes can vary a lot during the game and also from  game to another. 

\begin{figure}[!ht]
         \includegraphics[width=0.5\textwidth]{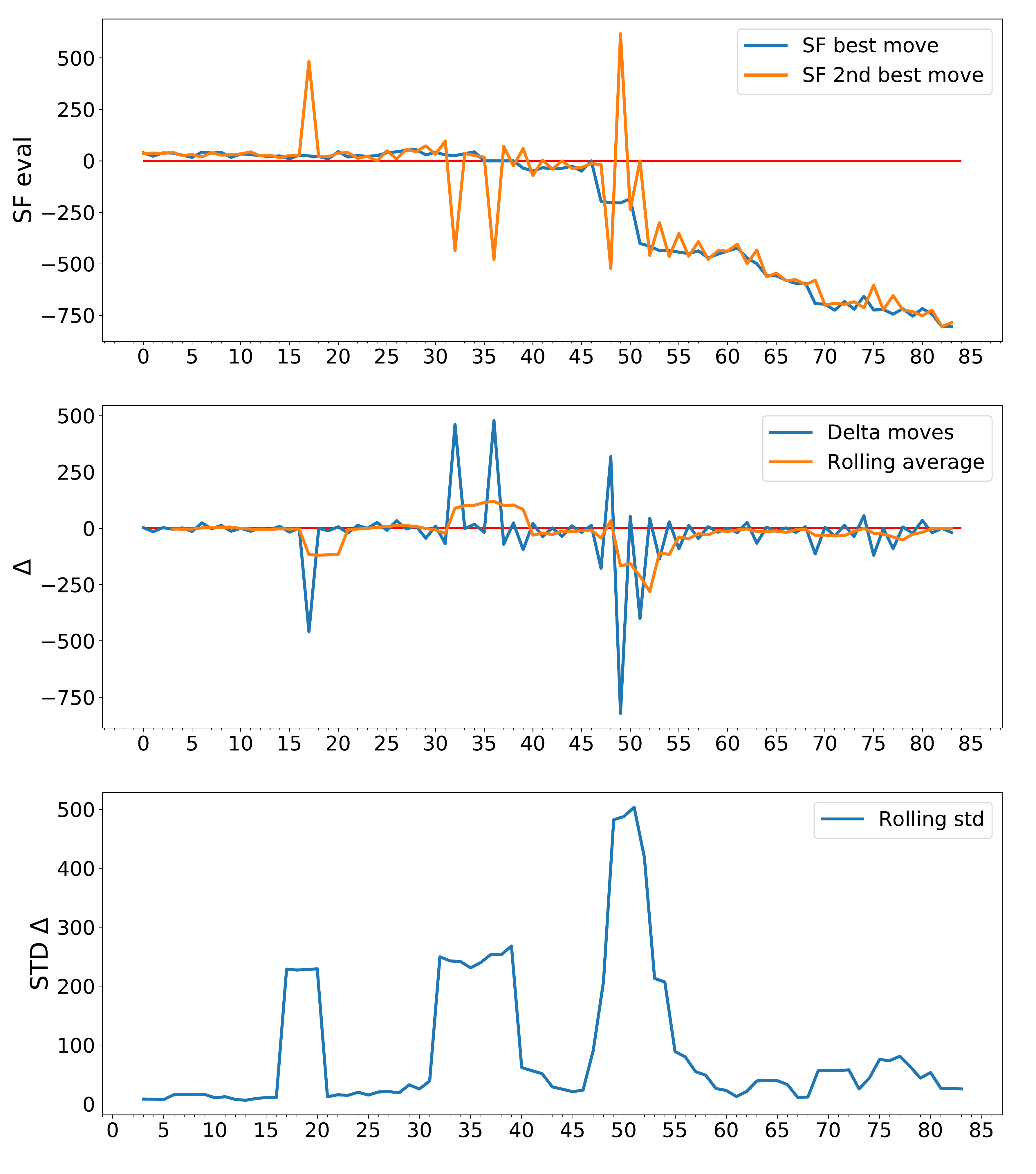}
         \caption{Calculation of $\Delta$ for the game between Benko and Keres (1963). (a) Stockfish evaluation for the best move and the second best move. (b) Evolution of $\Delta=E_0-E_1$ during the game (together with its rolling average computed over 4 half-moves). (c) Evolution of the Standard deviation (STD) of $\Delta$ during the game.} 
          \label{fig:delta2}
          \vspace{-5mm}
\end{figure}

 Another illustration is shown in Fig.~\ref{fig:delta2} where we plot results for the second round of the `First Piatigorsky Cup'  (Los Angeles, CA USA, 1963) between Pal Benko and Paul Keres (0-1). The first large value of $\Delta$ at ply $18$ is trivial: white captures the knight on \texttt{e4} and if black do not re-capture they loose a piece (which implies in general to loose the game). The next large $\Delta$ region appears at half move $32$ (for black): exchanging the bishop (with \texttt{Bxc3}) is by far the best move (with $\Delta\approx -100$). After this move, the large value for $\Delta$ corresponds again to a trivial move with the recapture of the bishop. The game is almost equal for both sides until half move $48$ which appears as a turning point in the game with the surprising move \texttt{e5g6} which is the best move with $\Delta=-140$ compared to queen's exchange (with \texttt{Qxe4}).  It is interesting to note that Dorfman \cite{Dorfman} discussed this part of the game as a critical point. After this point, we are in a regime of large $\Delta$ fluctuations but the game is now in favor of black.

\subsection{Distribution of $\Delta$}

\begin{figure}[!ht]
  \begin{center}
    \includegraphics[width=0.4\textwidth]{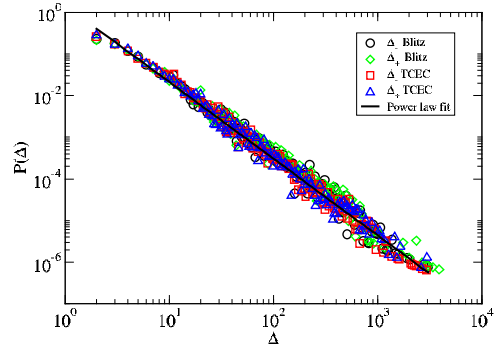}
  \end{center}
  \caption{Distribution of the quantity $\Delta$ (difference between the Stockfish evaluation for the best move and the second best move) computed for a sample of about $100$ blitz games \cite{data:blitz} and about $100$ games played by engines during the TCEC23 \cite{data_tcec23}.  We separate positive values ($\Delta_+$) from negative values ($\Delta_-$) and plot both distributions $P(\Delta_{\pm})$ in each case. The power law fit gives a value $\beta=1.84$ ($r^2=0.99$).} 
	\label{fig:powerlaw}
 \end{figure}

In order to explore the statistics of $\Delta$ and the presence of non-trivial tipping points, 
we exclude the exchange of pieces that are usually forced moves. After excluding these trivial
moves, we separate the positive values of $\Delta$ -which correspond to moves with white pieces - from the negative ones which correspond to moves with black pieces. We denote the corresponding values $\Delta_+$ and $\Delta_-$, respectively. We observe that $|\Delta|$ can vary over several order of magnitudes and display maximum values of order $10^3$ while typically large values are of order $10^2$. In order to characterize the statistics of $\Delta$, we construct the distributions $P(\Delta_{\pm})$ for about $100$ games played during the World Rapid and Blitz Championships \cite{data:blitz}) (Warsaw, 2021) and $100$ played by the best engines during the Top Chess Engine Championship 2023 (TCEC23) won by Stockfish. The results are shown in Fig.~\ref{fig:powerlaw}.

We observe that $P(\Delta_+)=P(\Delta_-)$ indicating a symmetry between white and black sides. In addition, we found that these distributions can be fitted by the same power law of the form
\begin{align}
  P(\Delta)\sim \frac{1}{\Delta^\beta}
\end{align}
with $\beta\approx 1.84$. In addition, this exponent seems to be `universal' and intrinsic to the game, as it doesn't seem to depend on the fact that players are human or engines.  The existence of such a power law indicates that during chess games, we can have regimes with very large fluctuations of $\Delta$ indicating a relatively high frequency of tipping points that determine in a large part the outcome of the game. This point certainly deserves further investigations, in particular to understand the emergence of such a power law and how we could determine the tipping points from the spatial organization of pieces.

\section{Discussion and outlook}

Maybe surprisingly, the scientific analysis of chess games is, for a long time now, a work in progress. Two major breakthrough however appeared these last decades that could accelerate our scientific understanding of this game. The first one is the emergence of engines that go far beyond human capabilities and brought some `fresh air' to the theory by revisiting some openings, and even in some cases denied commonly accepted principles \cite{Sadler:2019}. These engines can also be used for discovering strategical principles by analyzing many games. This brings us to the second important point: the availability of large databases with a huge number of games. The possibility to explore a large number of games opens the intriguing question of a statistical analysis of chess games. Finding patterns in these games might provide a way to creative strategical thinking.

In particular, we observed statistical differences between engines and human players. The probability to find a piece at a certain location is very different between humans and engines. More generally, we observe important differences in the use of pieces for the control of squares. In general, it seems that engines control space in a more uniform way (with a smaller gradient from the center to the borders), while a human player focuses on a smaller number of squares (central squares mainly). An important ingredient for achieving this is the structure of pawns and how they are used. Stockfish seems to move its pawns faster and further than most humans.  These algorithms seem then to optimize the spatial control with all pieces and do not focus on central squares. Obviously, this is here a first attempt to compare engines and human players, and more work is needed in this direction in order to understand fundamental strategical differences (if any) between human and engines. 

Finally, we observed tipping point that determine in a large part the outcome of the game. It would be crucial to understand the emergence of these tipping points and how to characterize them. In particular, it would be interesting to analyze these tipping points from the perspective of the spatial distribution of pieces. 

This paper is an attempt to provide a scientific approach to the chess game with an empirical basis. In particular, a statistical approach is somehow orthogonal to classical approaches done by chess players, but could point to interesting patterns and principles in this game. We provided here some tools and some results about chess games but the important point that we want to emphasize is the use of statistical methods (applied to a large number of games) for understanding strategical principles - beyond tactics and calculations that are necessary in any case. We believe that this approach could constitute an interesting starting point for a scientific approach to chess, and obviously many further interesting research remains to be done. Below, we discuss briefly some directions that could constitute directions for further studies.

\subsection{Pawn chain dynamics}

\begin{figure}[!ht]
	\begin{center}
		\includegraphics[width=0.5\textwidth]{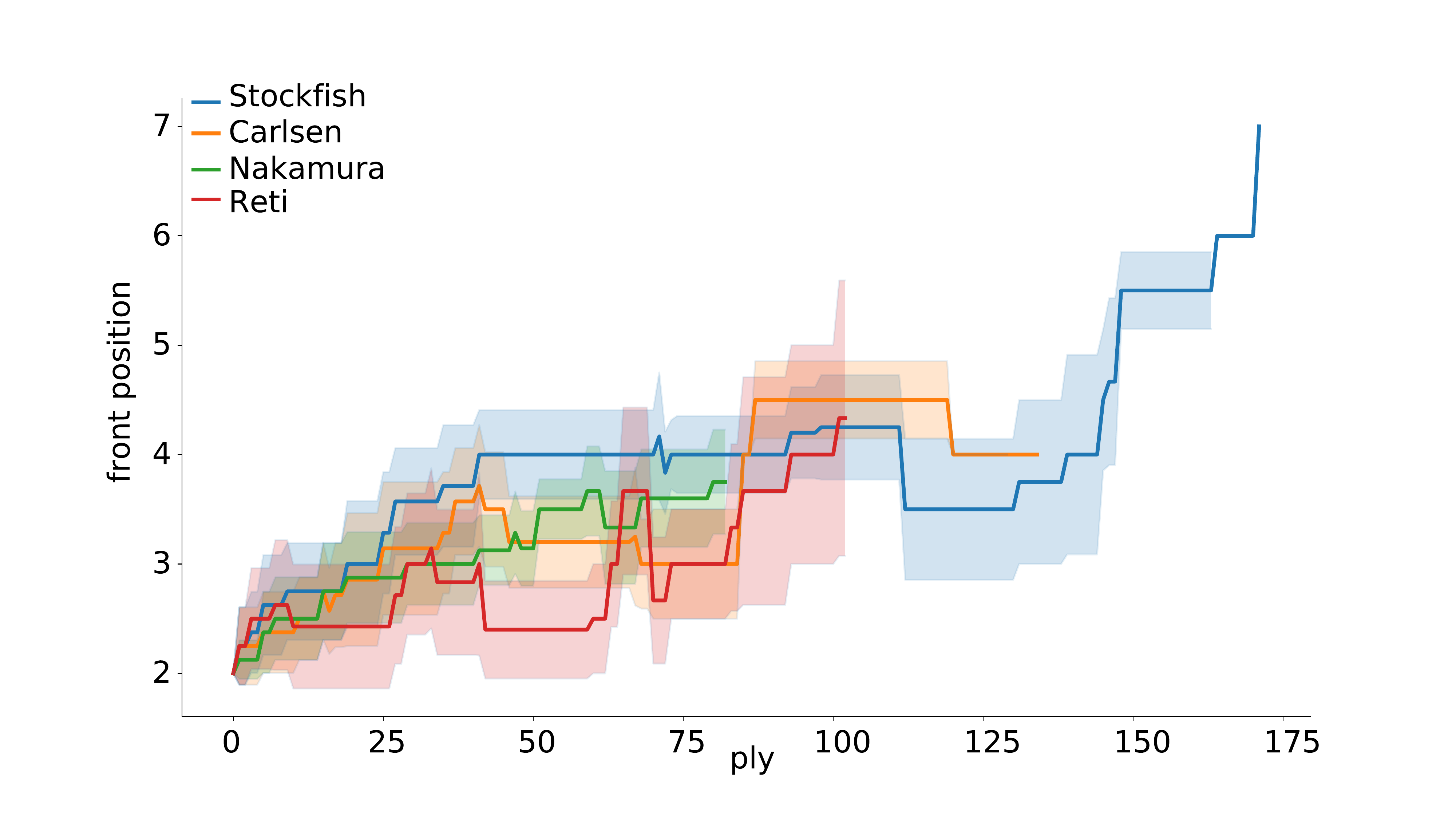}
	\end{center}
	\caption{Dynamics for the (white) front pawn for Stockfish15. Carlsen, Nakamura and Reti. We show here the average pawn front position versus the ply number. The shaded area corresponds to the average $\pm$ one standard deviation.} 
	\label{fig:front}
\end{figure}

According to the first world champion Wilhelm Steinitz, the pawn structure is the position's skeleton (see for example \cite{Grooten}). The structure of the pawn front indeed determines where the pieces can go and define the amount of space that is controlled by one side. This is a salient point in openings which consist essentially to build up a structure that will allow for the largest number of controlled squares and to eventually win the game. It is believed that a deep understanding of these pawn structure is a key ingredient for winning games. An interesting research direction could then be the quantitative study of the pawn structure. For example, the `pawn front' could
be described by a vector of $8$ elements $v=(v_1,v2,v_3,...,v_8)$ where each element denotes the position of the pawn in the corresponding column (from \texttt{a} to \texttt{h}). During the game, this vector evolves and we can characterize it by its average (i.e. the average position of the pawn front) and its dispersion (when one or more pawns are captured, we don't take them into account in the computation of these statistical properties). The evolution of the pawn front (for white) is shown for a few players in Fig.~\ref{fig:front}.
In general, we observe that the average front position is larger for Stockfish than human players, and with a smaller dispersion. If we had to use an image, engines such as Stockfish move most of their pawns and fast. At a smaller scale, the pawn front structure gives interesting information about players behaviors. We observed that Stockfish uses all its pawns and move then as far as possible, and that the best human player at this time - Magnus Carlsen - is following this trend. Obviously, a more detailed study (for different openings for example) is needed, but we believe that the pawn structure, at the heart of the chess game, certainly deserves further studies.

\subsection{Interaction graph}

\begin{figure}[!ht]
  \includegraphics[width=0.5\textwidth]{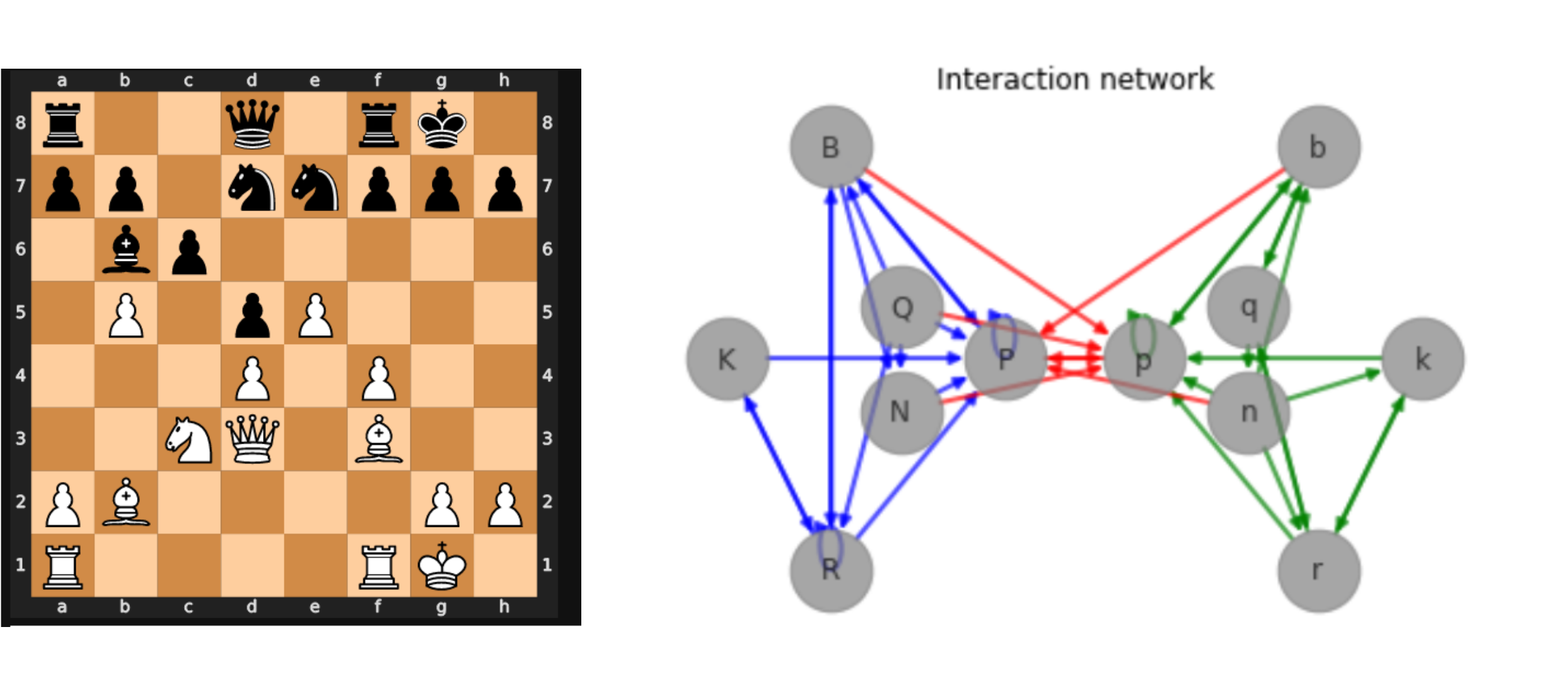}
	\caption{Example of a position and the corresponding interaction graph. The blue and green arrows represent the interaction between pieces of the same color (for example, the white queen protects the bishop on \texttt{f3}, the rook on \texttt{f1}, the knight on \texttt{c3}, and the pawn on \texttt{d4}, while the red arrows denote the attacks on the enemy (for example the bishop on \texttt{f3} threatens the \texttt{d4} pawn). Capital letters correspond to white pieces and lowercase to black pieces. } 
	\label{fig:graph}
\end{figure}

The number of controlled squares, the fluctuations of $\Delta$, etc. are important parameters
but we cannot ignore that pieces are not isolated and interact with each other. Pieces of the same color defend each other, sometimes in bad cases obstruct each other, and pieces of different colors can threaten each other. For pieces of the same color, we could analyze their interaction and look if there is a some sort of a `synergy' and how they attack the pieces of the opposite color. We can represent these interactions with a small graph where the nodes are the pieces and the links represent the `protection' when they have the same color, or the `attack' for different colors. We show an example of such an `interaction graph' in Fig.~\ref{fig:graph}.
This graph corresponds to the $30^{th}$ half move of a game between Savrov and Alekhine. Intuitively,we can expect that some properties of this graph are correlated with the quality of the position. The more the pieces protect each other, the better the position. This suggests
that the density of the subgraph for example (number of links/max number of links) could be a good indicator of the coherence or synergy of pieces of the same color. We could also ask ourselves how to characterize the quality of the position from the attacking point of view. Graphs (and more generally hypergraphs) are able to encode multi-body interactions
and this might be an interesting tool for discussing the quality of a position.

\subsection{Testing Dorfman's method}

The International Grand Master Iossif Dorfman proposed a move-search algorithm \cite{Dorfman}. The first point in this method is to detect what he calls `critical points'. According to Dorfman, these critical points correspond to either a position where an exchange of pieces is possible, or the pawn structure can be modified, or the end of a series of forced moves. Once, the critical points have been determined, one has then to assess the position based on various elements  described by Dorfman, and depending on this assessment the player has to make either a conservative move or in contrast a `dynamical' move that will modify strategical elements. 

In particular, the first point would be to test Dorfman's definition of critical points. The tipping points defined here might provide a way to identify Dorfman's critical points (a preliminary analysis shows that there is indeed a correlation between Dorfman's critical points and the tipping points defined as the onset of large $\Delta$ fluctuations). The next step would then be to test Dorfman's method at these critical points. This method that is well-known at the top-level - but didn't reach a consensus - would greatly benefit to the chess world if proven scientifically.

\section*{Acknowledgments}

This work could not have been done without Michael Bon and I thank him warmly for numerous discussions and for helping me to understand many things in Chess. I also thank Michael for having organized a discussion with the International Grand Master Iossif Dorfman. Finally, I also thank A. Flammini for interesting discussions.

\section{Methods}

\subsection{Data}

There are many available online resources for chess games. We used for our analysis exclusively open access data that can for example be found in \cite{data1, data2}. Typically for a given player, we have a large number of games (of order a few thousands) under the $\mathrm{.pgn}$ format which includes metadata about the game (date, location, opponent, etc), and the moves in chess algebraic notation (where the columns of the chessboard go from \texttt{a} to \texttt{h} and the rows from $1$ to $8$).  For each color (black and white) there are $6$ different pieces denoted by P=pawn, R=rook, N=knight, B=bishop, Q=queen, and K=king (and lowercase letters for black pieces).

We used data for particular openings obtained in \cite{data1} and also games from the World Rapid and Blitz Championships \cite{data:blitz}) that took place in Warsaw (2021) with top players (Magnus Carlsen, Hikaru Nakamura, Jan-Krzysztof Duda, Ian Nepomniachtchi, Fabiano Caruana, Anish Giri, Alexander Grischuk, Sergey Karjakin etc.). We also analyzed games played by the best engines  during the Top Chess Engine Championship 2023 (TCEC23) won by Stockfish \cite{data_tcec23}, and also some games played during the FIDE World Championship 2023 that took place in Astana (KAZ) the 2023.04.10 between Nepomniachtchi and Liren \cite{FIDE:2023}.

\subsection{Software}

We used python-chess (v1.9.4) which is a chess library for Python, with move generation, move validation, and support for common formats \cite{pythonchess}.

In particular, it implements an easy-to-use Stockfish class to integrates the Stockfish chess engine with Python \cite{pypisf}. We run all our simulations with depth ranging from $20$ to $30$.

\subsection{Games discussed here}

Games are generally recorded with the algebraic notation. This notation is based on a system of coordinates composed of a letter (from \texttt{a} to \texttt{h}) that indicates the position on a x-axis (from the point of view of white) and a number (from $1$ to $8$) for the y-coordinate. All games in the \texttt{pgn} format are recorded with this notation.

In addition to statistics computed over a large number of games, we discussed in particular three games  and we provide here all the corresponding moves. These games can for example be downloaded at \cite{data1} and \cite{FIDE:2023}).

\vspace{3mm}
\subsubsection{Benko-Keres, 1963}

The first game was played during the `First Piatigorsky Cup' (Los Angeles, CA USA) in 1963, between the two Grand masters Pal Benko and Paul Keres. The opening is here E15 (the Queen's Indian defense), and Keres won this game in 84 half moves.

1.c4 Nf6 2.Nf3 e6 3.d4 b6 4.g3 Ba6 5.Qa4 Be7 6.Nc3 Bb7 7.Bg2 O-O 8.O-O Ne4
9.Nxe4 Bxe4 10.Rd1 Qc8 11.Ne1 Bxg2 12.Nxg2 c5 13.d5 exd5 14.Rxd5 Nc6 15.
Bd2 Bf6 16.Bc3 Bxc3 17.bxc3 Re8 18.Ne3 Rxe3 19.fxe3 Qe8 20.Qc2 Qxe3+ 21.Kh1 Ne5 22.Rf1 Re8 23.Rf4 f6 24.Qe4 Ng6 25.Qxe3 
Rxe3 26.Rxd7 Nxf4 27.gxf4 Rxe2 28.Rxa7 Rf2 29.Rb7 Rxf4 30.Rxb6 Rxc4 31.Rb3
Kf7 32.Kg2 g5 33.Kf3 Ke6 34.Ra3 h5 35.Ke2 Rh4 36.Ra6+ Ke5 37.a4 c4 38.Rc6 
Rxh2+ 39.Ke3 Rh3+ 40.Kd2 Rd3+ 41.Kc2 h4 42.Rxc4 Rd8 0-1

\subsubsection{Liren-Nepomniachtchi, 2023}

The second game discussed here is between Liren and Nepomniachtchi, and is the second round of the FIDE World Championship 2023 that
took place in Astana (KAZ) the 2023.04.10. Nepomniachtchi won this game after 29 moves (all the games are freely available at \cite{FIDE:2023}).

1.d4 Nf6 2.c4 e6 3.Nf3 d5 4.h3 dxc4 5.e3 c5 6.Bxc4 a6
7. O-O Nc6 8. Nc3 b5 9. Bd3 Bb7 10. a4 b4 11. Ne4 Na5 12. Nxf6+
gxf6 13. e4 c4 14. Bc2 Qc7 15. Bd2 Rg8 16. Rc1 O-O-O 17. Bd3
Kb8 18. Re1 f5 19. Bc2 Nc6 20. Bg5 Rxg5 21. Nxg5 Nxd4 22. Qh5 f6
23. Nf3 Nxc2 24. Rxc2 Bxe4 25. Rd2 Bd6 26. Kh1 c3 27. bxc3 bxc3
28. Rd4 c2 29. Qh6 e5 0-1

\subsubsection{Mehedlishvili-Van Forrest, 2022}

The third game considered here is between Mehedlishvili and VanForrest during the Chennai Chess Olympiad in 2022 \cite{Chennai}.
Van Forrest won this game with black after 41 moves.

1. Nf3 d5 2. c4 d4 3. b4 c5 4. b5 Nf6 5. g3 Qc7 6. Bg2 e5 7. d3 Bd6
  8. O-O O-O 9. a4 Bf5 10. e4 dxe3 11. Bxe3 a5 12. Nc3 Nbd7 13. Nh4 Bg4
  14. Qc2 Rad8 15. h3 Be6 16. Nf3 h6 17. Nd2 Qc8 18. g4 Rfe8 19. Rad1 Nf8
  20. b6 Ng6 21. Nb5 h5 22. Ne4 Nxe4 23. dxe4 Bb8 24. f3 Rd4 25. Nxd4 exd4
  26. Bd2 hxg4 27. hxg4 Nf4 28. Bxa5 Qd8 29. Rfe1 Be5 30. Bf1 g6 31. Bd2 Kg7
  32. Bxf4 Bxf4 33. e5 Rh8 34. Re4 Qh4 35. Bg2 Be3+ 36. Rxe3 dxe3 37. Qe2 Qh2+
  38. Kf1 Qf4 39. Rc1 Rd8 40. Rc3 Qg3 41. Rc2 Bxc4 0-1


\end{document}